\documentclass[conference, letter]{IEEEtran}
\IEEEoverridecommandlockouts

\ifCLASSINFOpdf
  \usepackage[pdftex]{graphicx}
  \DeclareGraphicsExtensions{.pdf,.jpeg,.png}
\else
\fi

\usepackage{cite}
\usepackage{amsmath,amssymb,amsfonts}
\usepackage{algorithmic}
\usepackage{graphicx}
\usepackage{textcomp}
\usepackage[left=1.29cm, right=1.29cm, top=1.9cm, bottom=4.29cm]{geometry}
\usepackage{enumitem}
\usepackage{anyfontsize}
\usepackage{subcaption}
\usepackage{hyperref}
\usepackage{fancyhdr}
\makeatletter
\renewcommand{\fnum@figure}{Figure \thefigure}
\makeatother

\title{{Air Traffic Management for Collaborative Routing of Unmanned Aerial Vehicles via Potential Fields} \\
\thanks{This research has been financially supported by NASA University Leadership Initiative for the project "Autonomous Aerial Cargo Operations at Scale".}
}

\author{\IEEEauthorblockN{Josue N. Rivera, Dengfeng Sun}
\IEEEauthorblockA{School of Aeronautics and Astronautics \\
Purdue University \\
West Lafayette, IN, USA \\
\{river264, dsun\}@purdue.edu}
}

\renewcommand\IEEEkeywordsname{Keywords}
\IEEEaftertitletext{\vspace{-1\baselineskip}}

\begin{document}

\maketitle
\thispagestyle{fancy}

\noindent \begin{abstract}
Aerial cargo transport is anticipated to play a pivotal role in the distribution of goods within urban environments. The shift is propelled by the surge in e-commerce, the imperative to deliver essential supplies to isolated areas, and the growing demand for expedited and more accessible deliveries. Our research introduces a quantifiable standard for defining routing restrictions for Unmanned Aircraft System Traffic Management (UTM) using the concept of repulsive potential fields. Furthermore, we propose a scalable infrastructure that facilitates collaborative routing of cargo Unmanned Aerial Vehicles (UAVs) by independent shareholders. The practicality of the infrastructure is validated through a functional prototype implemented at a national scale. 
\end{abstract}

\vspace{0.3cm}

\begin{IEEEkeywords}
air traffic management; unmanned aircraft system traffic management; path planning; air transportation governance; information management; drone delivery
\end{IEEEkeywords}

\section{Introduction}

Aerial cargo delivery is anticipated to play a pivotal role in urban goods transportation. The surge in e-commerce over the past few years and the escalating demand for expedited delivery from customers have catalyzed the need for sustainable alternatives to traditional delivery methods \cite{boysen2021last, morganti2014impact}. The demand for cargo unmanned Aerial Vehicles (UAVs) to deliver critical medicines and other essential goods to remote areas remains as another significant driving force \cite{scott2017drone, rabta2018drone}. However, regulatory and privacy concerns present substantial obstacles to the evolution of aerial delivery systems \cite{sah2021analysis}. It is imperative to establish an infrastructure that is scalable and adaptable to a diverse set of policy requirements \cite{ telli2023comprehensive, shrestha2021survey, sah2021analysis}.

A primary motivation for this research is to translate such policies into quantifiable metrics. Metrics that can establish limitations for cargo UAVs and assist in formulating routing solutions for Unmanned Aircraft System Traffic Management (UTM). As such, our work introduces a standard for defining these restrictions and proposes an air traffic management infrastructure that takes advantage of it. The infrastructure would facilitate the deployment of a collaborative platform, enabling independent parties involved in aerial cargo deliveries to operate within the same airspace. This research primarily focuses on last-mile cargo delivery, where UAVs transport packages within an urban environment.

\subsection{Background}
    
Distinct approaches have been proposed for a UTM infrastructure that can support aerial cargo deliveries operations. One such approach is the design of a hybrid truck-drone model \cite{madani2022hybrid, wang2020cooperative, huang2022solving, wang2019routing}. The design primarily consists of equipping delivery trucks with the capability to launch UAVs that can deliver packages and returns to their vehicle as the trucks travels through an urban space. A few limitations arise from the approach. The coordination of such system would require solving complex variances of the well-known traveling salesmen problem at a large scale for diverse environments \cite{agatz2018optimization}. The optimization problem may prove to be a challenging task in a scalable infrastructure. A different popular approach is the use of aerial corridors \cite{wang2022route, sacharny2020large, rastgoftar2023finite, nagrare2023multi}. It primarily consists of defining an aerial network of predefined paths through which cargo UAVs can travel to their destination. The infrastructure can be made scalable, however, the rigid fly paths may lead to inefficient routes and may not be feasible for the limited range of some UAVs. Additionally, variances of the approach do not take advantage of the often unobstructed airspace.
    
To address some of the limitations with the designs of these UTM infrastructures, we propose the use of potential fields. Artificial repulsive potential fields are a concept that treats agents as particles moving within a space, influenced by repulsive forces emitted from obstacles \cite{hwang1992potential}. Potential fields are commonly used in path planning and the development of routing algorithms for ground vehicles and robots moving in obstructed spaces  \cite{dubey2023path, aggarwal2020path}. In the realm of traffic management, repulsive artificial potential fields have also been employed to manage coordination of multiple agents. \cite{han2022potential, hu2023potential}, among others, have successfully explored them for non-aerial traffic management systems and demonstrated their success as a traffic management tool. Our work distinguishes itself by being one among a selected few exploring potential fields in aerial routing and presenting a scalable UTM infrastructure that relies on them.

Recently, \cite{jana2023numerical, tang2019optimized} proposed the use of potential fields for UTM. These approaches present UTM infrastructures that consider moving UAVs or high-density areas as obstacles within a 3D potential field. They demonstrate collision-free collaborative aerial routing in various urban scenarios. Our work expands upon the ideas presented by the authors in three significant ways. First, we standardize the creation of potential fields to ensure a scalable approach. Second, we simplify the problem to enhance adaptability to distinct urban environments by considering aerial routing on a 2D plane. Lastly, we extend the discussion on the applicability of potential fields in existing low-altitude airspace and their use in collaborative aerial cargo deliveries by independent parties.

\subsection{Assumptions and Context}

\begin{figure}[h]
  \centering
  \includegraphics[width=0.3\textwidth]{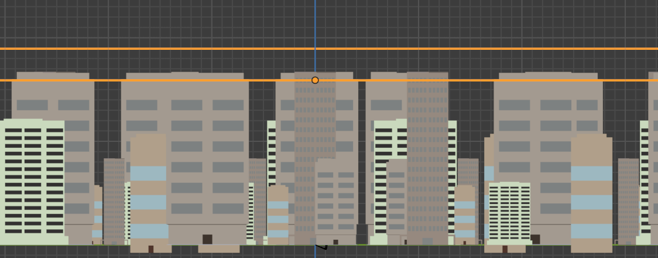}
    \caption{Aerial cargo delivery operating airspace sector in a hypothetical city. The operating height would ideally be above most buildings and having minimal interaction with the airspace of low-flying aircraft.}
    \label{fig:flight-sector}
\end{figure}

In our study, we make the assumption that cargo UAVs operate at a constant altitude (e.g., 500+ feet) from the ground and in low-altitude airspace. This altitude forms a plane for a potential field that associates intercepting buildings and other constraints as restrictions. The research primarily focuses on unidirectional cargo delivery trips. The UAVs are assumed to be equipped with Vertical Take-Off and Landing (VTOL) capabilities, or alternatively, they possess a mechanism that enables precise pick-up and drop-off of cargo at their destination and origin. Although the ideal operating height would be above most buildings, the presence of unavoidable structures is anticipated. The shared airspace with other low-altitude aircraft is also factored into consideration. Fig. \ref{fig:flight-sector} provides a visualization of the operating airspace sector.

\subsection{Contributions}

Our work brings forth the following main contributions:

\begin{itemize}
  \item A standard for defining Urban Air Mobility (UAM) routing restrictions.
  \item An infrastructure for scalable and quantifiable UTM of cargo UAVs.
  \item A functional prototype of the proposed infrastructure.
\end{itemize}
\section{Potential Field and Geographical Restriction}

\begin{figure}[!b]
  \centering
    \begin{subfigure}[b]{0.23\textwidth}
        \centering
        \includegraphics[width=\textwidth]{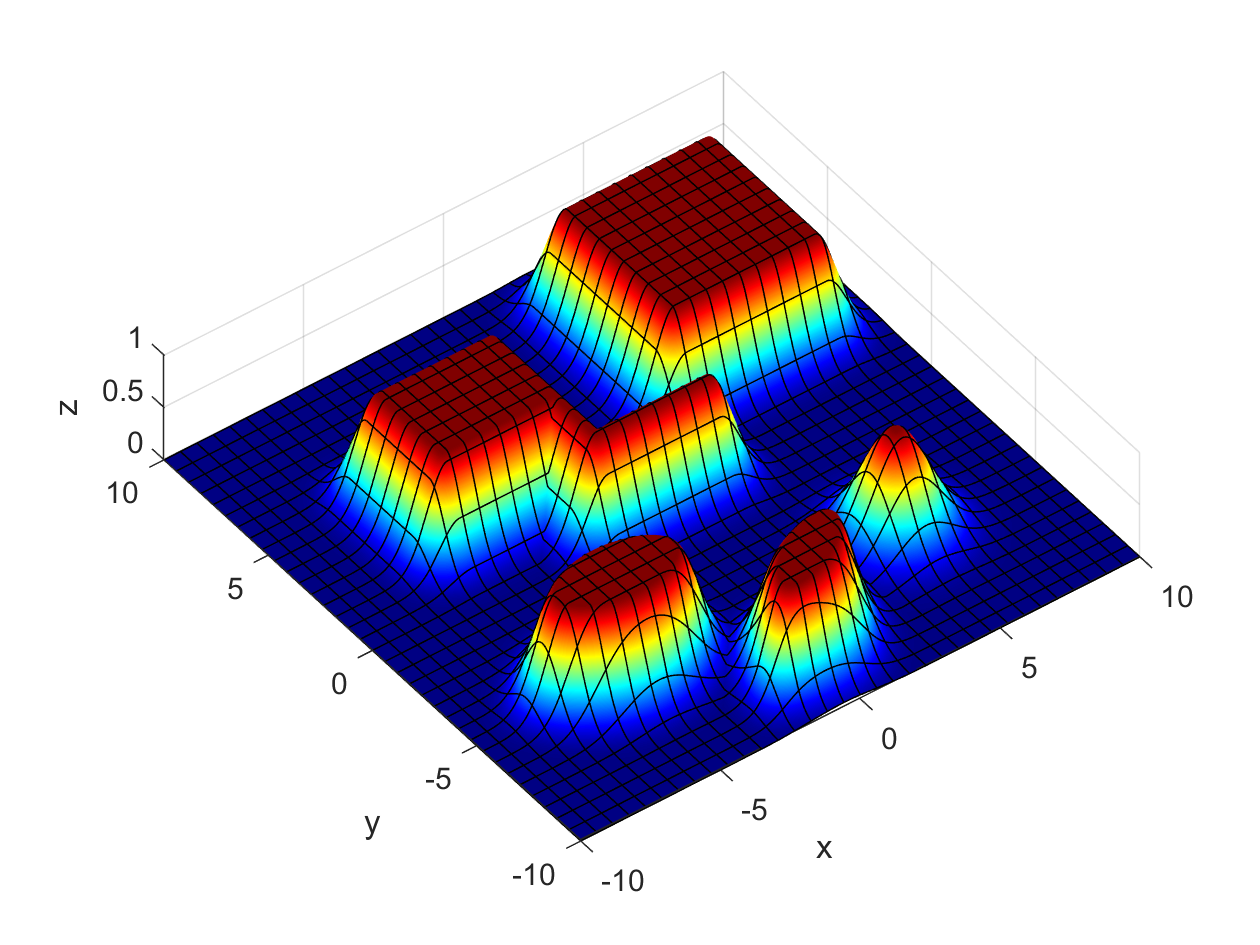}
        \caption{Potential field viewed from an angle.}
    \end{subfigure}~
    \begin{subfigure}[b]{0.23\textwidth}
        \centering
        \includegraphics[width=\textwidth]{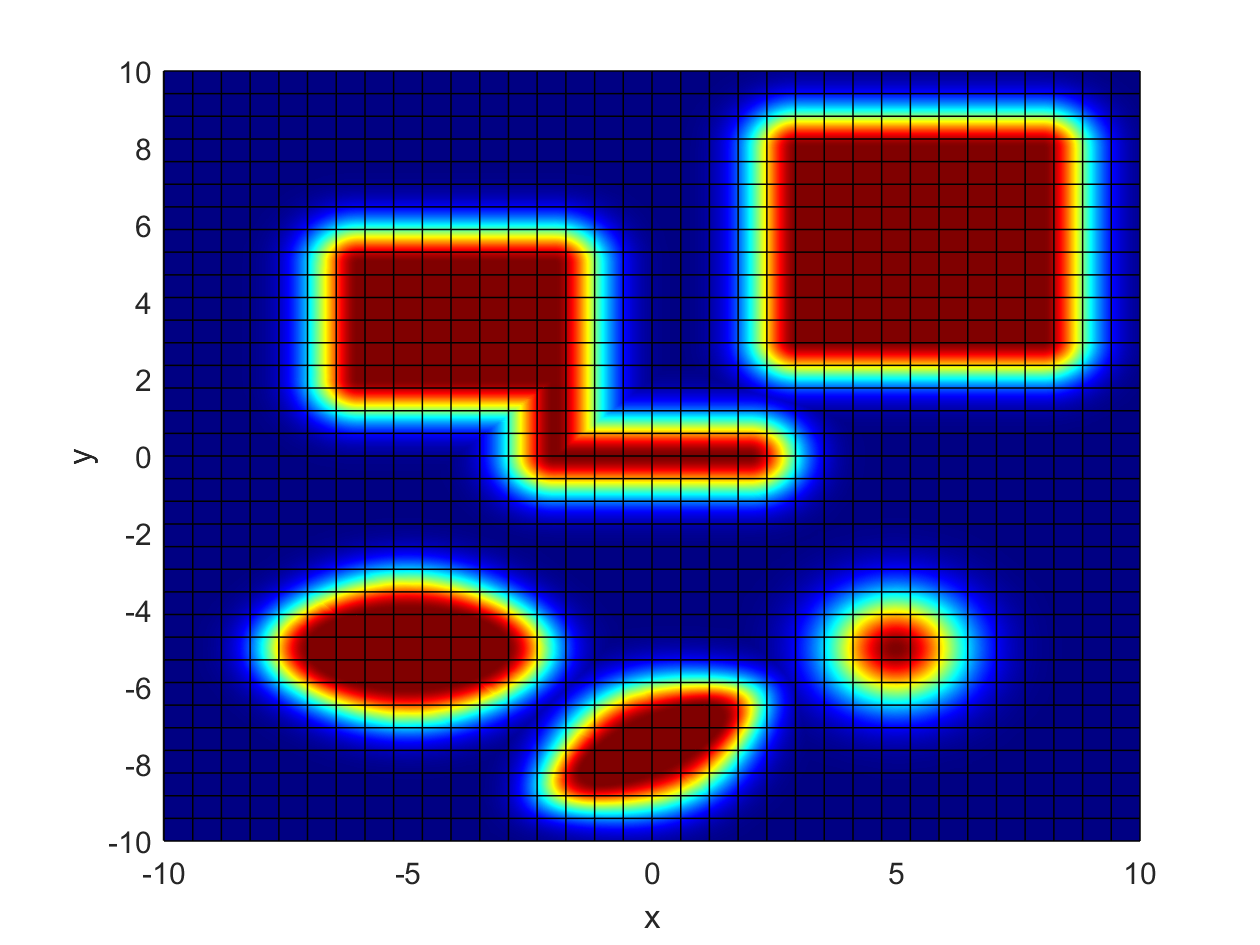}
        \caption{Potential field viewed from above.}
    \end{subfigure}
    \begin{subfigure}[b]{0.23\textwidth}
        \centering
        \includegraphics[width=\textwidth]{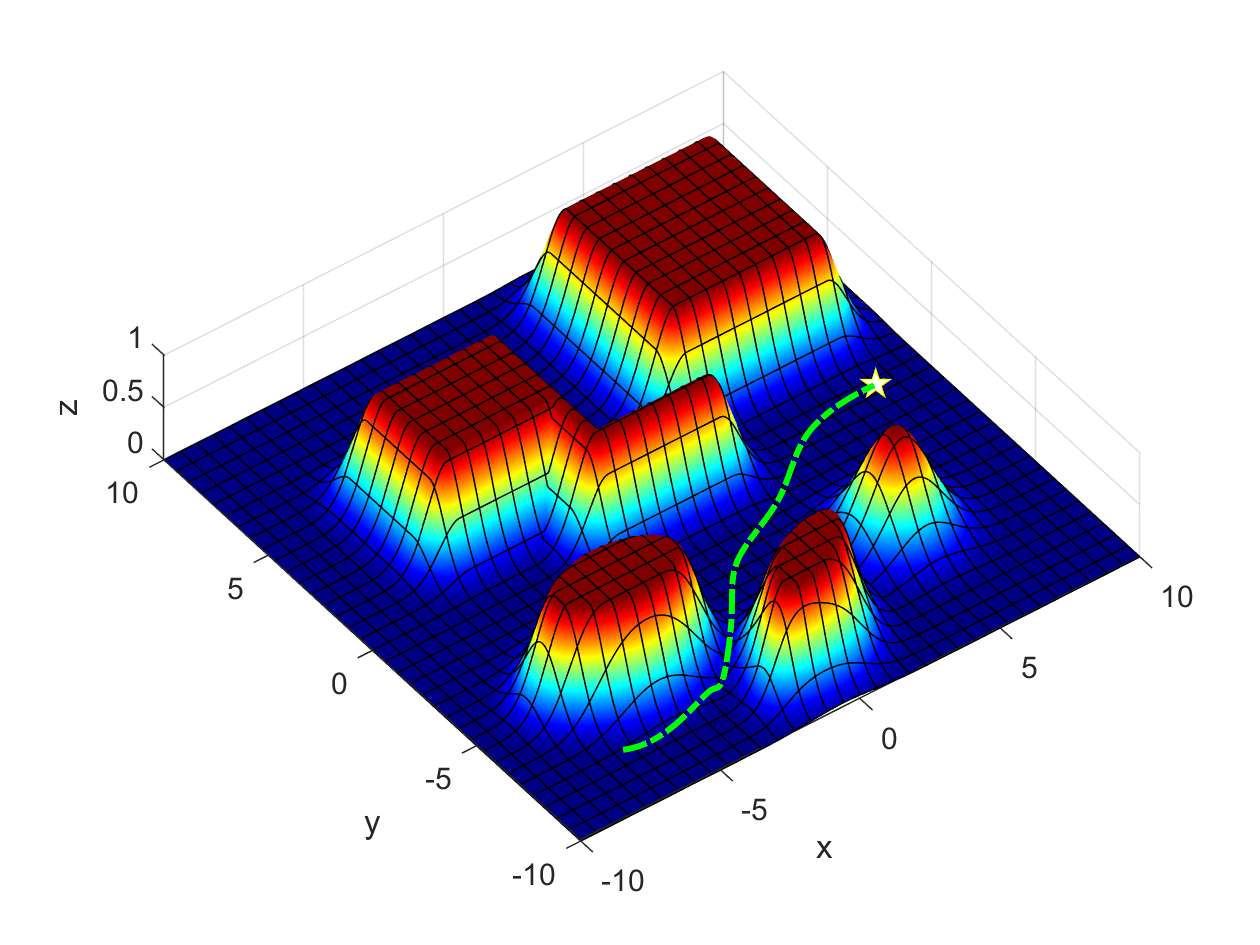}
        \caption{Potential field with a route, viewed from an angle.}
    \end{subfigure}~
    \begin{subfigure}[b]{0.23\textwidth}
        \centering
        \includegraphics[width=\textwidth]{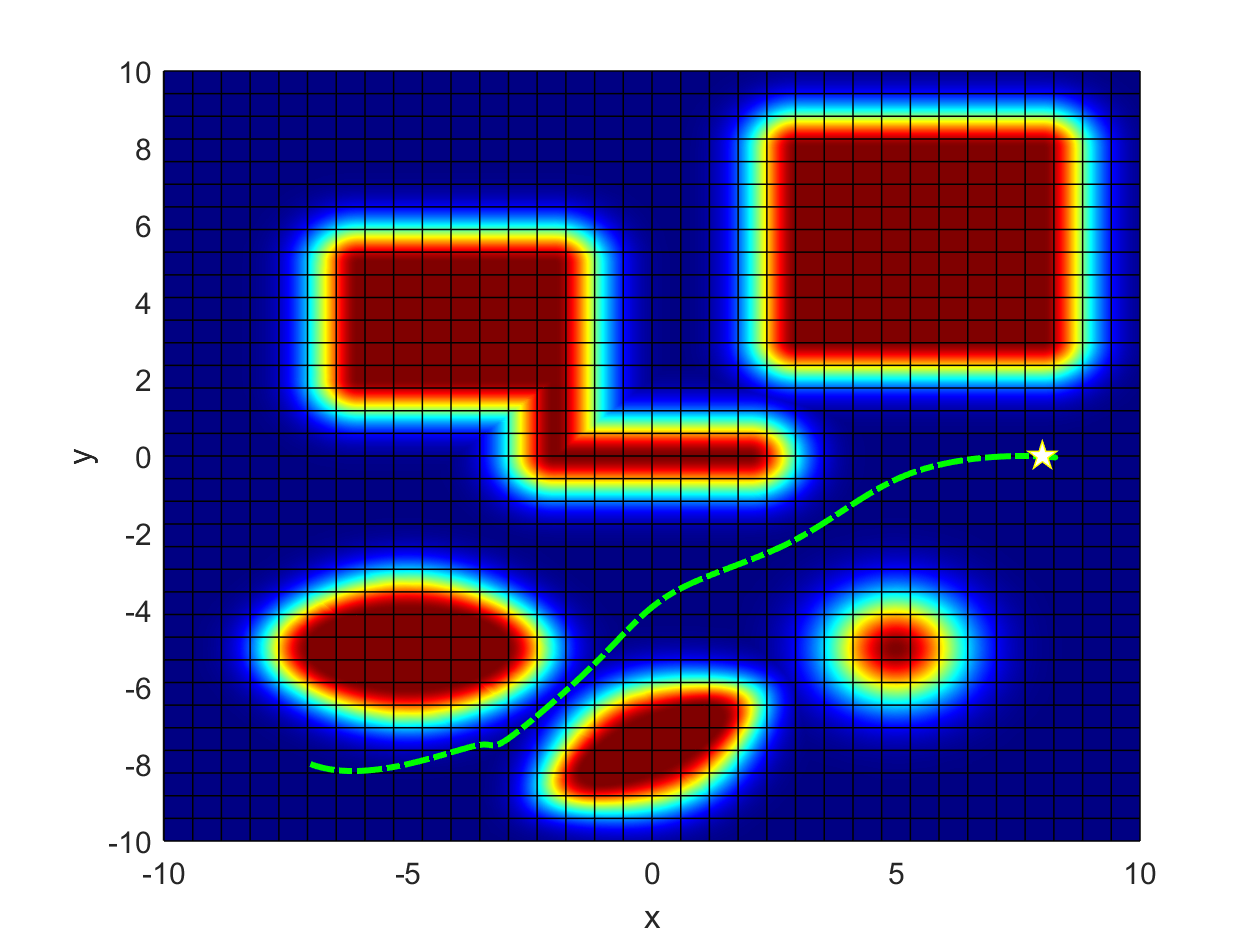}
        \caption{Potential field with a route, viewed from above.}
    \end{subfigure}
    \caption{3D surface plot of a potential field and a low-energy route.}
    \label{fig:field-mix}
\end{figure}

To define UAM routing restrictions, we propose a standardized repulsive potential field as a quantifiable representation of the limitations. The field reflects geographical constraints and the desired behavior of UAVs in their vicinity. It associates regions to avoid with a high energy state, while all other locations are associated with an energy value that decays the further they are from these regions. In this proposed standard, an energy value of 1 signifies a complete restriction, while a value of 0 indicates complete freedom for flight. The potential field aims to represent the diverse requirements of stakeholders in an unified format.  Fig. \ref{fig:field-mix} illustrates an example of a potential field and a route that has a minimal area under its path.

\subsection{Fundamental Units for Potential Field}
\label{sect:field-units}

\begin{figure*}
  \centering
    \begin{subfigure}[b]{0.24\textwidth}
        \centering
        \includegraphics[width=\textwidth]{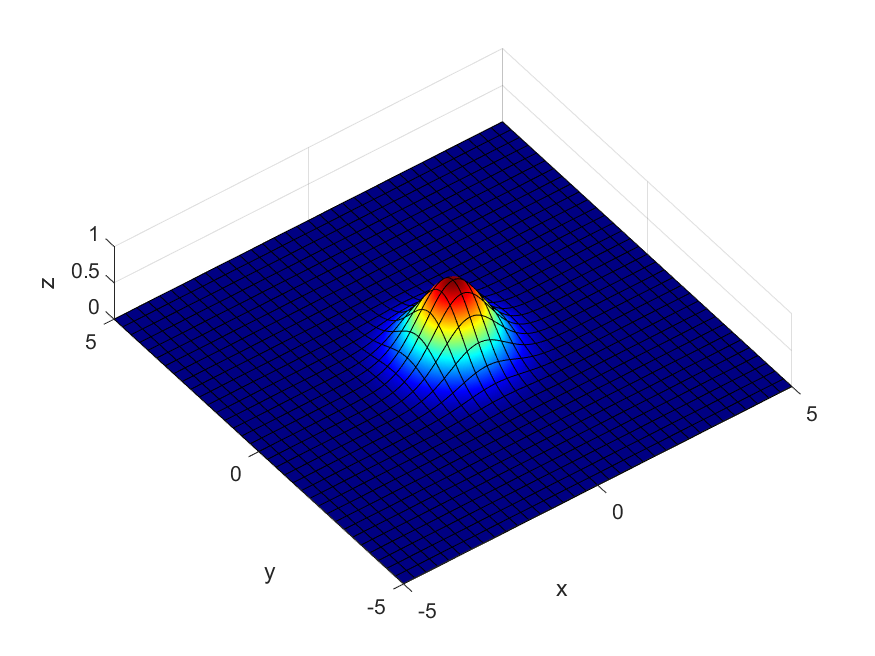}
        \caption{Point unit}
        \label{fig:sub:point-field}
    \end{subfigure}~
    \begin{subfigure}[b]{0.24\textwidth}
        \centering
        \includegraphics[width=\textwidth]{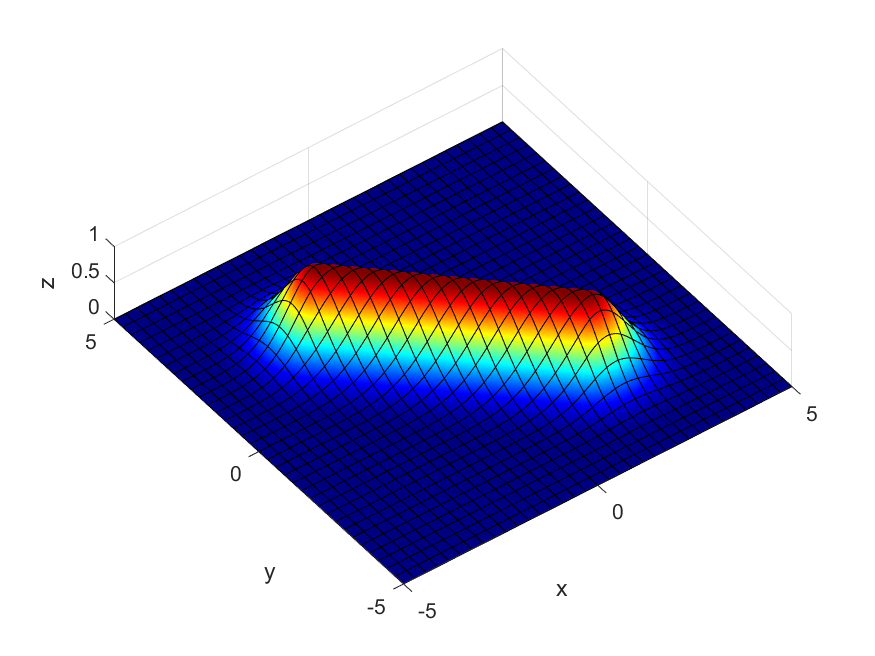}
        \caption{Line unit}
        \label{fig:sub:line-field}
    \end{subfigure}~
    \begin{subfigure}[b]{0.24\textwidth}
        \centering
        \includegraphics[width=\textwidth]{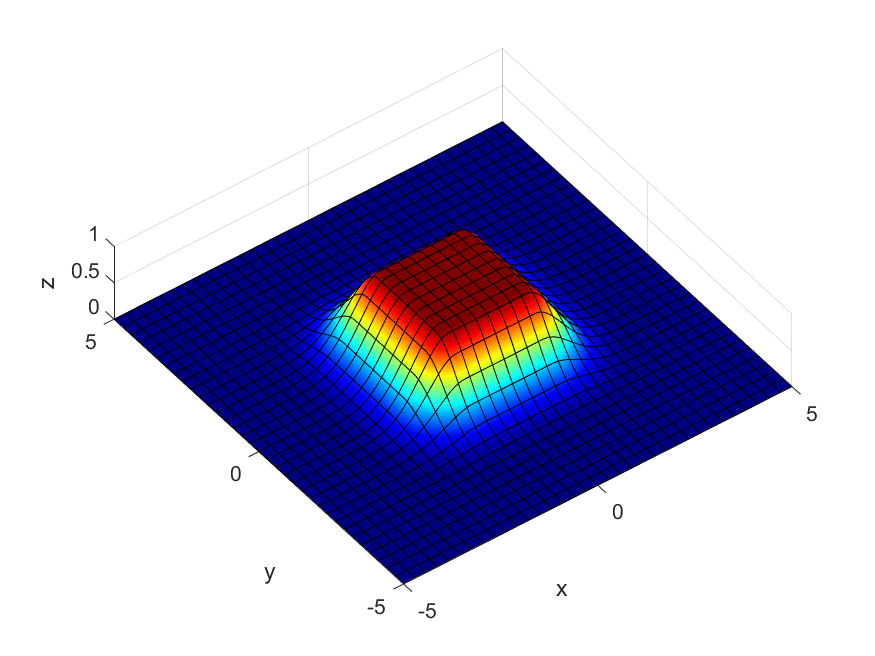}
        \caption{Rectangle unit}
        \label{fig:sub:rect-field}
    \end{subfigure}~
    \begin{subfigure}[b]{0.24\textwidth}
        \centering
        \includegraphics[width=\textwidth]{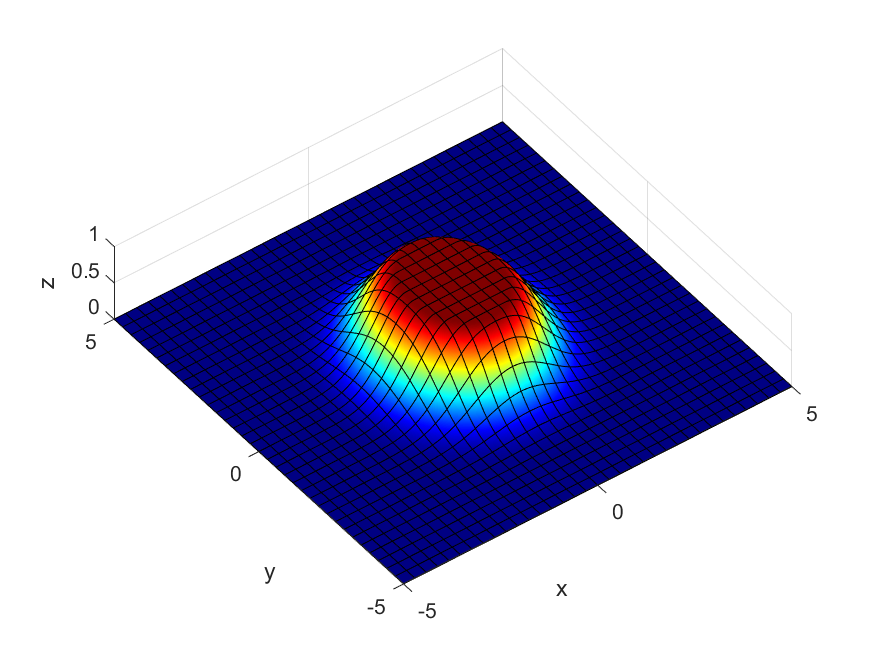}
        \caption{Ellipse unit}
        \label{fig:sub:ellipsoid-field}
    \end{subfigure}
    \caption{Fundamental units that form the basis of complex potential fields.}
    \label{fig:fundamental-units}
\end{figure*}

The standardized potential field is built from smaller fundamental units. The units $\sigma_i$ represent individual physical or virtual restrictions in an urban environment that affect the travel of cargo UAVs. Each distinct unit has a set of parameters that defines their impact in the potential field such as where it is placed and the shape of the repulsive force. Examples of the potential field for these fundamental units are presented in Fig. \ref{fig:fundamental-units}. These units may be integrated to form complex fields $\Omega$ via a maximum operator (or a variance of it). 

\begin{equation}
    \Omega(x) = \max \left (\sigma_1(x|\cdot,\mathrm{A}_1),\, \sigma_2(x|\cdot,\mathrm{A}_2),\, ...,\, \sigma_N(x|\cdot,\mathrm{A}_N) \right)
    \label{eq:combined-field}
\end{equation}

In \eqref{eq:combined-field}, $x$ represents a given point to evaluate, and $\mathrm{A}_i$ a positive definite matrix that describes the repulsion from a high energy state. It should be noted that all potential field units have a maximum and minimum energy value of 1 and 0, respectively, to maintain a normalized scale.

\subsubsection{Point} 

The most basic potential field unit is the point. It highlights a spot in a map to avoid and is described by,

\begin{equation}
    \sigma_p(x|\hat{x},\mathrm{A})=e^{-{(x-\hat{x})}^T\mathrm{A}^{-1}(x-\hat{x})}
\end{equation}
where $x$ is an evaluated point, $\hat{x}$ is the unit location, and $\mathrm{A}$ is a positive definite matrix that expresses the repulsive force.

As a point is impractical as a barrier, the potential field is intended to discourage travel through a given area without applying a strict restriction. The potential field can be view in Fig. \ref{fig:sub:point-field}.

\subsubsection{Line}

The potential field unit for a line is defined by three parameters. Two points, $\hat{x}_1$ and $\hat{x}_2$, indicate the start and end of a line, respectively. The positive definite matrix $\mathrm{A}$, as with a point, expresses the repulsive force. The process of determining the potential field unit begins by finding the point on the line that is closest to a given point x. This is achieved by,

\begin{equation}
    \begin{split}
    P_l\left(x\middle|\hat{x}_1,\hat{x}_2\right)&=\frac{(\hat{x}_2-\hat{x}_1)\cdot(x-\hat{x}_1)}{\|\hat{x}_2-\hat{x}_1\|^2}\\
    g_l\left(x\middle|\hat{x}_1,\hat{x}_2\right)&=\hat{x}_1+\mathop{\rm clamp}\left(P_l\left(x\middle|\hat{x}_1,\hat{x}_2\right) ,0,1\right)(\hat{x}_2-\hat{x}_1)
    \end{split}
    \label{eq:point-unit}
\end{equation}
where $\mathop{\rm clamp}(\alpha, a, b)$ is a function that limits a given value $\alpha$ to the range $a$ to $b$. Following this, the point $x$ is then evaluated with respect to new point $\hat{x} = g_l\left(x\middle|\hat{x}_1,\hat{x}_2\right)$.

\begin{equation}
    \sigma_l\left(x\middle|\hat{x}_1,\hat{x}_2,\mathrm{A}\right)=e^{-{(x-g_l\left(x\middle|\cdot\right))}^T\mathrm{A}^{-1}(x-g_l\left(x\middle|\cdot\right))}
\end{equation}

The potential field for a line can be used to restrict flights along paths in a map. Paths can often be simplified into a set of points with segmented lines connected in-between. For instance, roads, rivers, and highways can be marked for avoidance by generating strings of segmented lines along their paths. An example of the potential field can be view in Fig. \ref{fig:sub:line-field}.

\subsubsection{Rectangle}

The potential field unit for a rectangle is more restrictive than that of a point or line. It defines a rectangular sector by $\hat{x}_1$ and $\hat{x}_2$ where any point $x$ inside it is classified as a high energy state. To define the field, we must first describe a new vector that signals the repulsion from the restricted area by,

\begin{equation}
    \begin{split}
    \bar{x}_e\left(x\middle|\hat{x}_1,\hat{x}_2\right) &= \frac{1}{2} \left(|x - \hat{x}_1|+|x - \hat{x}_2| - |\hat{x}_1 - \hat{x}_2|\right )\\
    g_r\left(x\middle|\hat{x}_1,\hat{x}_2\right) &= \mathop{\rm sign} (x - \hat{x}_1) \odot \bar{x}_e \left(x\middle|\hat{x}_1,\hat{x}_2\right)
    \end{split}
\end{equation}
where $|x|$ and $\mathop{\rm sign}(x)$ are the element-wise absolute value and sign operations, respectively, of the vector $x$. This new vector is then used to describe the potential field unit for a rectangle.

\begin{equation}
    \sigma_r\left(x\middle|\hat{x}_1,\hat{x}_2,\mathrm{A}\right)=e^{-g_r\left(x\middle|\cdot\right)^T\mathrm{A}^{-1}g_r\left(x\middle|\cdot\right)}
\end{equation}

The rectangle potential field unit is designed to restrict entities that can be accurately represented by rectangular shapes. These entities can include structures such as buildings, tracts of land, and other physical entities. Fig. \ref{fig:sub:rect-field} showcases an example of the potential field.

\subsubsection{Ellipse}
\label{sect:ellipse-unit}

The ellipse serves as an alternative to a rectangle. Given the shape of an ellipse, we define a new vector that signals the repulsion from the restricted area by,

\begin{equation}
    g_e\left(x\middle|\hat{x},\mathrm{B}\right)= \mathop{\rm max} \left(1 - \frac{1}{\|\mathrm{B}^{-1} \left(x-\hat{x}\right)\|}, 0\right) \left(x-\hat{x}\right)
\end{equation}
where $\hat{x}$ is the center of the ellipse, and $B$ is a positive definite matrix that describes its shape. As with a rectangle, this new vector is then used to describe the potential field unit.

\begin{equation}
    \sigma_e\left(x\middle|\hat{x},\mathrm{A},\mathrm{B}\right)=e^{-g_e\left(x\middle|\cdot\right)^T\mathrm{A}^{-1}g_e\left(x\middle|\cdot\right)}
\end{equation}

The ellipse unit is designed for restricting entities that possess, or can be more accurately estimated by, ellipsoidal shapes. These entities can include structures such as mountains, domes, and alike. A potential field for an ellipsoidal restriction can be seen in Fig. \ref{fig:sub:ellipsoid-field}.

\subsection{Virtual and Time-Dependent Potential Fields}

Potential fields are not necessarily tied to physical entities or restrictions at or above the operating altitude. They can be used to fully restrict or discourage UAV flights across chosen sectors. For example, they can deter or prevent flights across residential areas, parks, airports, runways, venues, open fields, and/or forests.

Moreover, potential fields, while not necessarily linked to physical entities, do not need to be static either. They can impose temporary restrictions around agents in the airspace, such as cargo UAVs and other low-altitude aircraft. They can also vary over time in response to certain virtual entities, such as shifts in population density throughout the day, or to discourage the use of similar flights within a short time frame.

\subsection{Geographical Restriction Encoding}

To encode and store the restrictions that form the basis of potential fields units, we propose a variant of the well-known GeoJSON format \cite{butler2016geojson}. GeoJSON, based on the JSON file format, is an open standard designed for representing simple geographical features and storing their non-spatial attributes. These features include points (representing addresses and locations), line strings (describing streets and boundaries), polygons (defining tracts of land and counties), and multi-polygons. Hence, to maintain interoperability, we introduce Restriction GeoJSON (RGeoJSON). 

\begin{figure}
  \centering
  \includegraphics[width=0.2\textwidth]{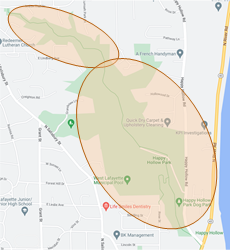}
    \caption{GeoJSON polygon geometry object of the park can be estimated by multiple RGeoJSON rectangles and/or ellipses. When compared to a single rectangle covering the park, the combined ellipses minimize the extent of the airspace that is restricted. Map courtesy of Google Maps.}
    \label{fig:polygon-ellipses}
\end{figure}

RGeoJSON brings three significant changes to the standard. First, it amends geometry objects to include properties for the repulsion matrix $\mathrm{A}$ (explored in Section \ref{sect:field-units}) and the matrix defining an ellipse shape $\mathrm{B}$ (discussed in Section \ref{sect:ellipse-unit}). Secondly, it removes support for polygon geometry objects. Lastly, it introduces two novel geometry objects: \textit{Rectangle} and \textit{Ellipse}. The addition of these objects compensates for the removal of GeoJSON polygons. Polygon shapes may be estimated by the use of multiple rectangles and ellipses as demonstrated in Fig. \ref{fig:polygon-ellipses}. 

As a variation, standard No-SQL databases that support GeoJSON offer some functionalities for RGeoJSON, such as geospatial queries for shared geometry object types. In Appendix \ref{sect:appendix}, a comprehensive list and examples of all RGeoJSON geometry objects are provided.
\section{Urban Air Mobility Infrastructure}

\begin{figure}
  \centering
  \includegraphics[width=0.45\textwidth]{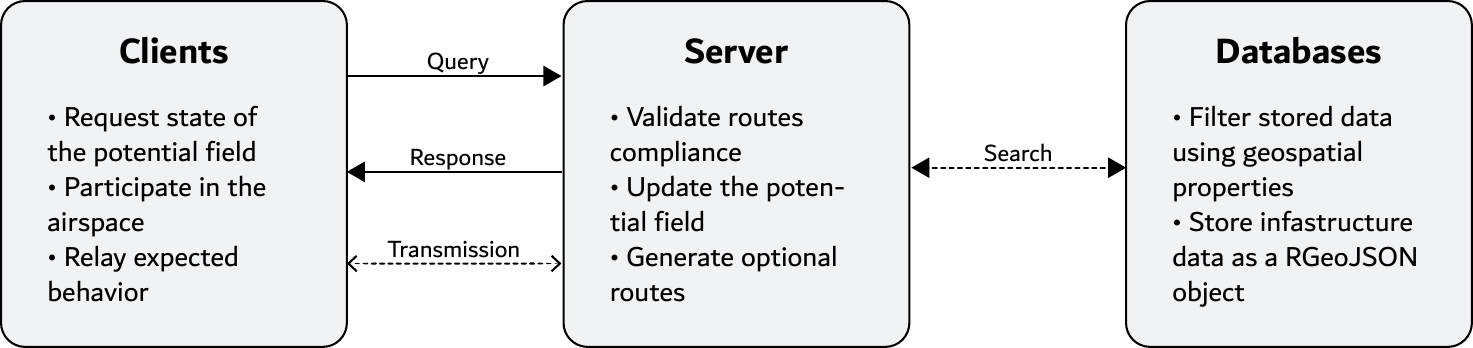}
    \caption{Overview of the proposed infrastructure and information exchange.}
    \label{fig:infrastructure-overview}
\end{figure}

In addition to the standard for defining UAM routing restrictions using potential fields, we propose an infrastructure to manage the airspace occupied by UAVs engaged in aerial cargo deliveries. The UTM infrastructure’s role is to facilitate the routing collaboration among independent third parties and to update the shared potential field as needed. This infrastructure consists of clients, a back-end server with distributed nodes, and databases for geographical restrictions. An overview of the infrastructure is presented in Fig. \ref{fig:infrastructure-overview}.

\subsection{Clients}

Clients in the infrastructure represent software and/or hardware that are participants or stakeholders in the airspace. These can be cargo UAVs, low-altitude manned vehicles, supply chain applications, among other systems. Each client has the ability to observe the state of the shared potential field and request updates to it. However, the level of authority and the type of requests a client can make are determined by the clearance they are assigned. Any physical client participating in the airspace must periodically transmit its current state and expected near-future behavior. It is the responsibility of the back-end server's governing body to define the interface through which clients can communicate.
        
\subsection{Server}

The back-end server acts as the governing body of the infrastructure, managing the coordination of independent third parties seeking to influence the shared airspace. It has three primary functions: generating routes given an origin and destination, validating clients’ requested routes, and managing the shared potential field for collision avoidance and desired policies.

When generating optional routes, the server applies a path-finding algorithm that uses the potential field to determine a low-energy route while minimizing its distance. However, when clients provide their own routes, the server shifts to evaluating the route compliance instead. This is achieved by calculating, or approximating, the area under the route in the potential field. To capture violations of the restriction, the field may first be transformed in such a way that it scales towards infinity near an energy value of 1, ensuring that any restriction violation results in an infinite value by the area approximation. In terms of managing the shared potential field, the server may introduce temporary changes to ensure coordination. One such change, similar to \cite{tang2019optimized}, involves requesting the transmission of the current and near-future locations of clients participating in the airspace and developing temporary potential fields around them. This would ensure some level of collision avoidance for any participating UAV.

To accommodate the scale of the shared potential field, the server may operate as a distributed system in which different nodes to support distinct geographical regions.

\subsection{Geographical Restriction Databases}

The geographical restriction databases stores infrastructure information of all locations of interest using the RGeoJSON format. These databases store the parameters needed to build the individual potential field unit while the back-end server decides how to serve them to the clients. As opposed to a client's requested changes, restrictions stored in the databases server are not considered temporary. These restrictions may be static or time-dependent for local regions. They may also either attached to physical or virtual obstacles to ensure safe aerial routing. Additionally, the databases may contain non-spatial information of the system, such as the operating altitude of given regions.
        
\subsection{Deployment} 

The UTM infrastructure is designed to enable communication between any participant in the airspace and the server. The server holds the responsibility of assigning authority and/or clearance to clients for the types of queries they are allowed make. As the central governing unit, the back-end server’s administration and operations must be overseen by an agency with the authority to manage airspace (e.g., Federal Aviation Administration or FAA). We recommend a distributed deployment of geographical restriction databases. This approach allows the back-end server to request data selectively, focusing only on the sectors relevant to the third-party operations. For practicality, counties and local municipalities with capable information technology (IT) departments are encouraged to host and maintain these databases. Such close integration would facilitate some governance of low-altitude airspace and ensure up-to-date information on restrictions affecting cargo UAVs. When a client submits a potential field state request, the server gets to consider the restrictions before providing a response.
\section{Implementation}

As a proof of concept, a prototype of the proposed infrastructure was implemented. It showcases the aforementioned components: a client, a server, and a geographical database.

\subsection{Potential Field State Client}

\begin{figure}[!b]
  \centering
    \begin{subfigure}[b]{0.235\textwidth}
        \centering
        \includegraphics[width=\textwidth]{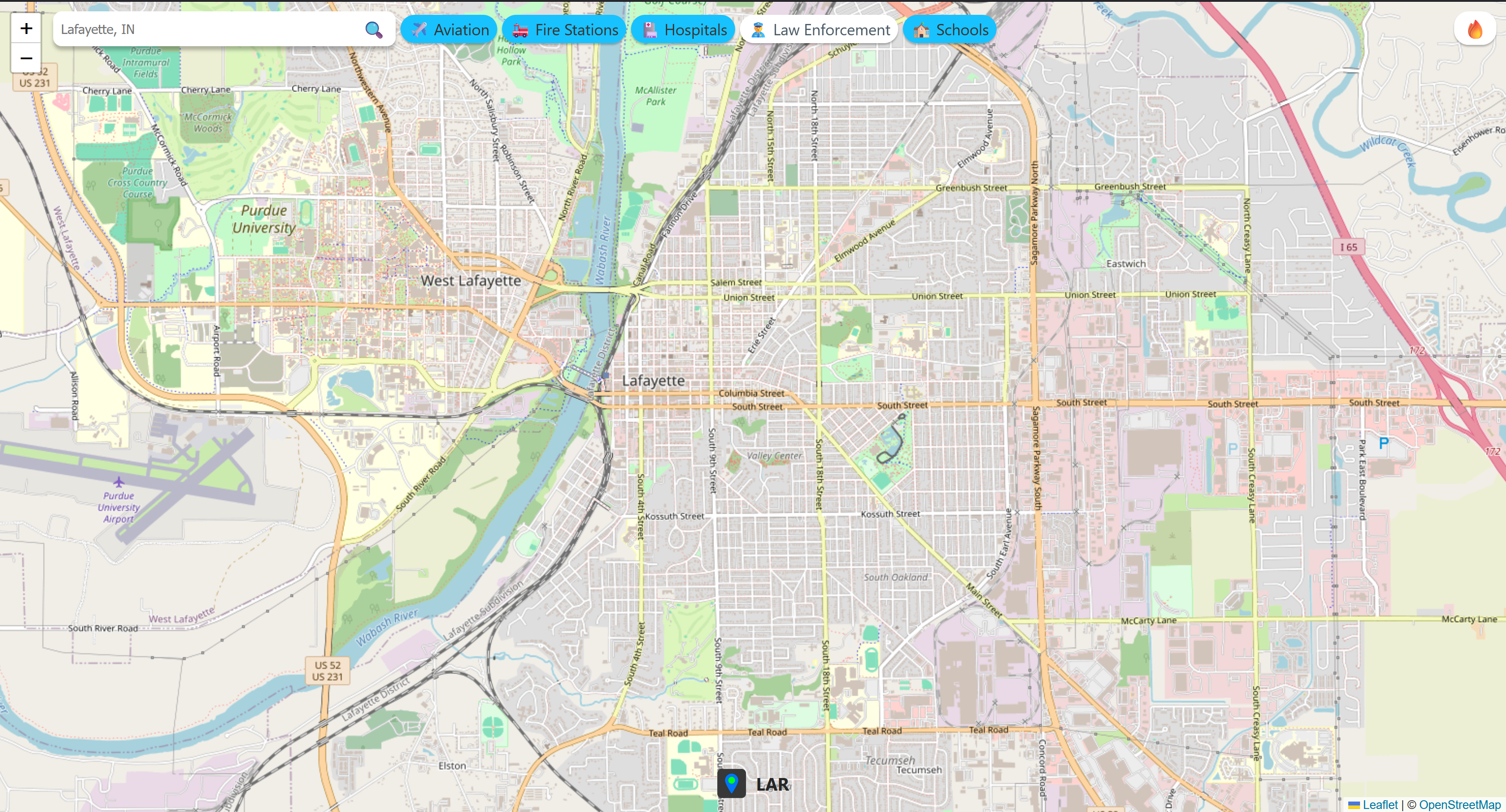}
        \caption{Map without the potential field state.}
    \end{subfigure}~
    \begin{subfigure}[b]{0.235\textwidth}
        \centering
        \includegraphics[width=\textwidth]{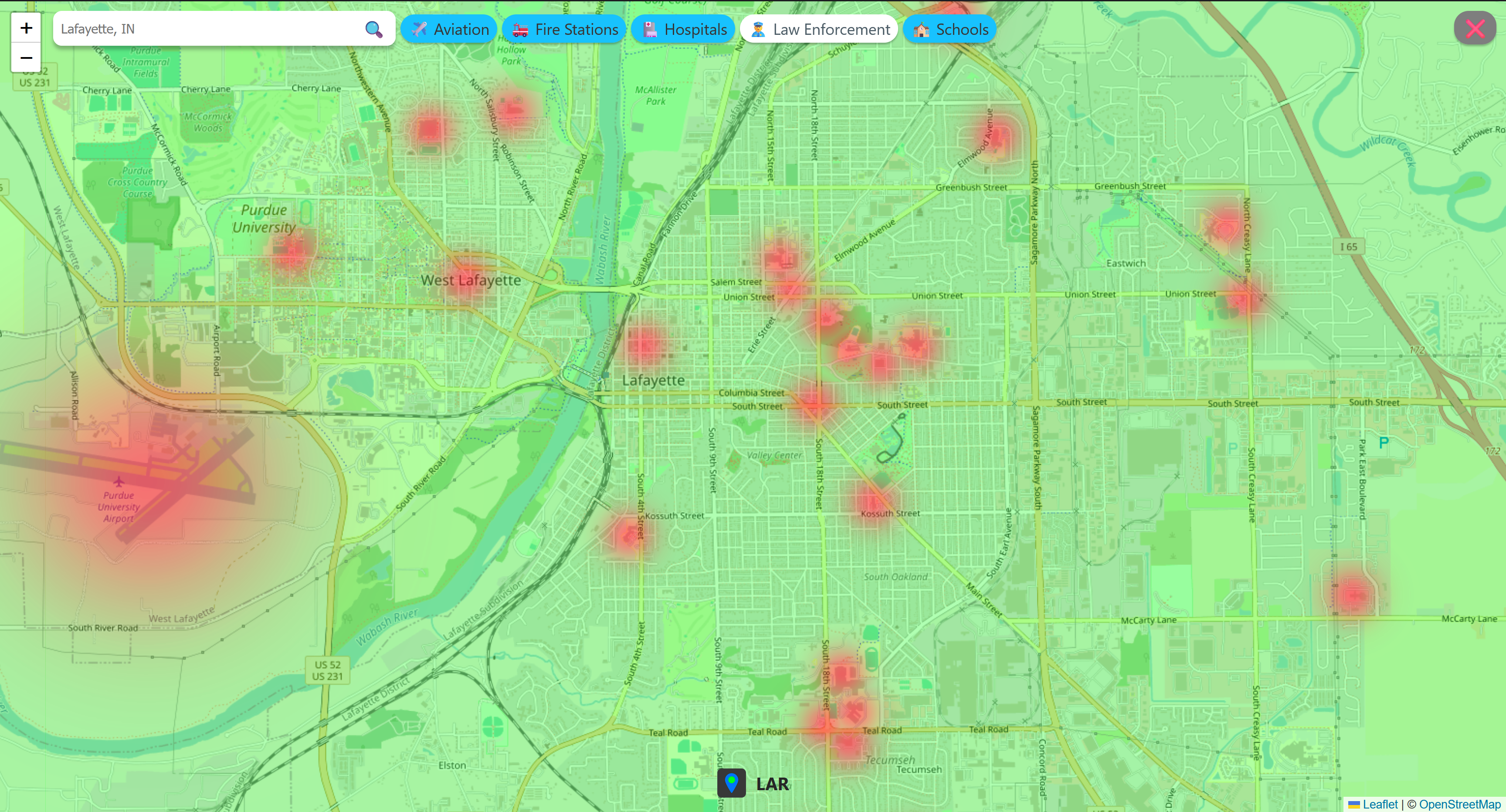}
        \caption{Map with the potential field state.}
    \end{subfigure}
    
    \begin{subfigure}[b]{0.48\textwidth}
        \centering
        \includegraphics[width=\textwidth]{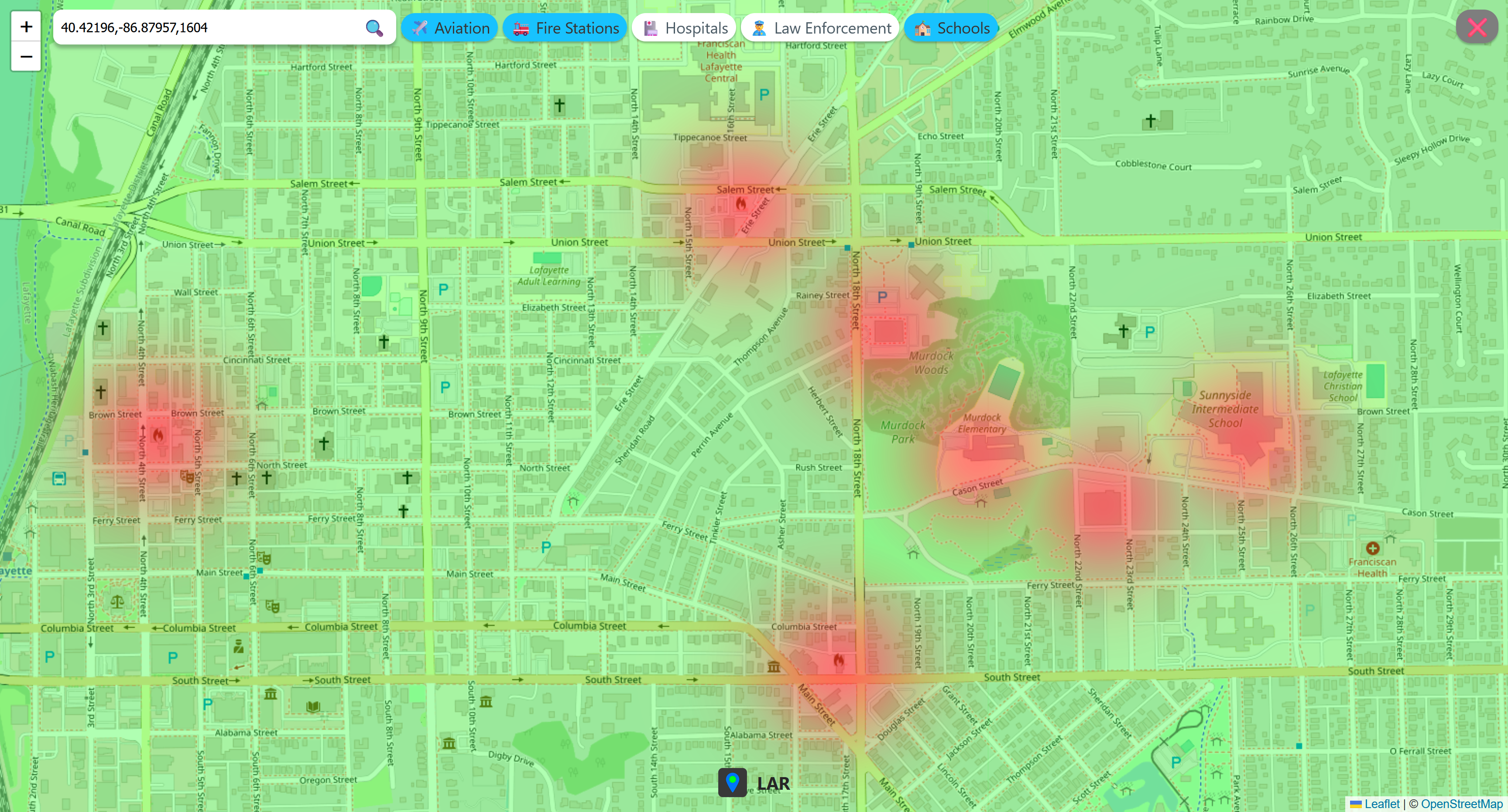}
        \caption{Zoomed map with the potential field state.}
    \end{subfigure}
    \caption{The potential field state of Lafayette, Indiana, USA is displayed by a React application operating as a client of the UTM infrastructure. For demonstration purposes, aviation facilities, fire stations, and schools are marked for restriction. These marked areas exhibit a diminishing energy value as the distance from them increases. The potential field is visualized using a translucent gradient that transitions from red to green.}
    \label{fig:frontend-app}
\end{figure}

The implemented client, serving as a front-end web server, projects the state of the potential field for any queried location within the United States. It is designed as a React application that dispatches HTTPS requests to the back-end server and utilizes the returned data to construct the potential field for the specified area. The potential field is then overlaid on the corresponding map region using a translucent filter. Fig. \ref{fig:frontend-app} showcases the user interface of the application.

The source code and documentation for the client are accessible on Github at \href{https://github.com/wzjoriv/LAR-frontend}{https://github.com/wzjoriv/LAR-frontend}.
        
\subsection{Back-End Server}

The implemented back-end server's software was developed using the Python language and its Flask package. Its primary function is to listen for client queries, create a thread to serve them, and then return to standby mode awaiting for the next client. In this infrastructure, HTTPS requests serve as the main communication interface, influencing the state of the potential field. For demonstration purposes, we implemented three primary HTTPS REST API queries:

\begin{enumerate}
    \item \textbf{GET /test/:adds} confirms the authority and clearance to communicate with the server for the provided address.
    \item \textbf{GET /adds/:adds/:dbs} returns the parameters defining the potential field for the provided address and infrastructure database collections.
    \item \textbf{GET /locs/:lat,:lon,:radius/:dbs} returns the parameters defining the potential field for the provided latitude, longitude, radius, and infrastructure database collections.
\end{enumerate}

While not indicative of a fully deployed UTM at a large scale, for this prototype, clients are permitted to submit a list of infrastructures for consideration. Both the list of infrastructures and a geospatial bounding restriction are utilized to search the geographical restrictions databases for the RGeoJSON Objects within the specified region.

Details on the server implementation and the source code can be found on Github at \href{https://github.com/wzjoriv/LAR-backend}{https://github.com/wzjoriv/LAR-backend}. 
        
\subsection{Database}

The geographical restrictions are stored in a MongoDB No-SQL database using the aforementioned RGeoJSON format. Among other features, MongoDB supports geospatial queries to filter results by their geographical location. If the server software used for this demonstration is connected to a MongoDB database, it will automatically download the presented dataset and upload it the database if it does not already exist. The dataset is an augmented version of multiple infrastructure datasets collected by the Homeland Infrastructure Foundation-Level Data (HIFLD) repository\footnote{HIFLD homepage: \href{https://hifld-geoplatform.opendata.arcgis.com/}{https://hifld-geoplatform.opendata.arcgis.com/}}. The datasets for US-based aviation facilities, fire stations, hospitals, law enforcement, and public schools were used. Geospatial search queries for any location within the United States is supported, but not all locations may have appropriate restriction parameters due to lack of data. 
\section{Conclusion}

In this paper, we proposed a standard for defining routing restrictions in UAM, coupled with a UTM infrastructure that leverages it for the operation of collaborative and scalable aerial cargo deliveries. The applicability of the infrastructure is demonstrated by a functional prototype that coordinates a shared potential field and operates at a national scale. 

The proposed infrastructure addresses regulatory concerns by: (1) operating in a manner that allows for governance of low-altitude airspace, and (2) offering functionalities to validate the compliance of independent third-party routes through a quantifiable approach. This work also seeks to mitigate some of the shortcomings of other UTM approaches by considering the scalability of the infrastructure and striving to avoid restrictions to predefined paths.

\subsection{Limitations}

While our research provides valuable insights, it is not without its limitations. One such limitation is the assumption of a near-constant operating altitude. Under this assumption, the search space of feasible routes outside of the operating altitude is unaccounted for, leading to sub-optimal routes. The assumption is a trade-off made to facilitate an adaptable solution to a wide range of urban environments and simplify the complexity of the problem. The limitation may be mitigated by the introduction of multiple potential fields at different altitudes and allowing UAVs to change layers depending on their departure and destination points. However, the implications of having multiple potential field layers in the proposed infrastructure falls outside the scope of this paper.

\subsection{Future works}

The research opens up several avenues for further exploration. One potential direction involves the development of optimized routing strategies that take full advantage of the newly introduced standard. For example, the repulsive gradient of the fundamental potential field units could be utilized to enhance the efficiency of route searches. Another promising avenue is the execution of complex, large-scale routing simulations for major cities. Such simulations could provide deeper insights into the practicality of implementing the proposed infrastructure within an urban environment. In a similar vein, investigating the impact of implementing multiple potential field layers at various altitudes is warranted. Lastly, with respect to the infrastructure and shared potential field, the real-time adaptation to changing conditions presents a compelling opportunity for future research.

\section*{Acknowledgment}

We extend our gratitude to Pruthvi S. Patel\footnote{Email: pate1533@purdue.edu} for his valuable contributions to this research. Patel assisted in various technical aspects of the project and supported its development.

\bibliographystyle{unsrt}
\bibliography{references}

\appendix

\subsection{RGeoJSON Geometry Object Examples}
\label{sect:appendix}

Following a conversion akin to GeoJSON, we present examples of UAM routing restrictions encoded using the RGeoJSON format \cite{butler2016geojson}. These examples illustrate the schema of the geometry objects used in the routing restrictions data storage.

\vspace{10pt}

\subsubsection{Point}
\begin{small}
\begin{verbatim}
{ 
    "type": "Point",
    "coordinates": [100.0, 0.0],
    "repulsion": [
        [25.0, 0.0],
        [0.0, 25.0]
    ]
}
\end{verbatim}
\end{small}

\subsubsection{LineString}
\begin{small}
\begin{verbatim}
{
    "type": "LineString",
    "coordinates": [
        [100.0, 0.0],
        [101.0, 1.0]
    ],
    "repulsion": [
        [25.0, 0.0],
        [0.0, 25.0]
    ]
}
\end{verbatim}
\end{small}

\subsubsection{Rectangle}
\begin{small}
\begin{verbatim}
{
    "type": "Rectangle",
    "coordinates": [
        [100.0, 0.0],
        [101.0, 1.0]
    ],
    "repulsion": [
        [25.0, 0.0],
        [0.0, 25.0]
    ]
}
\end{verbatim}
\end{small}

\subsubsection{Ellipse}
\begin{small}
\begin{verbatim}
{ 
    "type": "Ellipse",
    "coordinates": [100.0, 0.0],
    "repulsion": [
        [25.0, 0.0],
        [0.0, 25.0]
    ],
    "shape": [
        [50.0, 0.0],
        [0.0, 50.0]
    ]
}
\end{verbatim}
\end{small}

\subsubsection{MultiPoint}
\begin{small}
\begin{verbatim}
{
    "type": "MultiPoint",
    "coordinates": [
       [100.0, 0.0],
       [101.0, 1.0]
    ],
    "repulsion": [
        [25.0, 0.0],
        [0.0, 25.0]
    ]
}
\end{verbatim}
\end{small}

\subsubsection{MultiLineString}
\begin{small}
\begin{verbatim}
{
   "type": "MultiLineString",
   "coordinates": [
        [
           [100.0, 0.0],
           [101.0, 1.0]
        ],
        [
           [102.0, 2.0],
           [103.0, 3.0]
        ]
    ],
    "repulsion": [
        [25.0, 0.0],
        [0.0, 25.0]
    ]
}
\end{verbatim}
\end{small}

\subsubsection{MultiRectangle}
\begin{small}
\begin{verbatim}
{
    "type": "MultiRectangle",
    "coordinates": [
        [
           [100.0, 0.0],
           [101.0, 1.0]
        ],
        [
           [102.0, 2.0],
           [103.0, 3.0]
        ]
    ],
    "repulsion": [
        [25.0, 0.0],
        [0.0, 25.0]
    ]
}
\end{verbatim}
\end{small}

\subsubsection{MultiEllipse}
\begin{small}
\begin{verbatim}
{
    "type": "MultiEllipse",
    "coordinates": [
        [100.0, 0.0],
        [101.0, 1.0]
    ],
    "repulsion": [
        [25.0, 0.0],
        [0.0, 25.0]
    ],
    "shape": [
       [
            [50.0, 0.0],
            [0.0, 50.0]
       ],
       [
            [50.0, 50.0],
            [ 0.0, 50.0]
       ]
   ]
}
\end{verbatim}
\end{small}

\subsubsection{GeometryCollection}
\begin{small}
\begin{verbatim}
{
   "type": "GeometryCollection",
   "geometries": [{
        "type": "Point",
        "coordinates": [100.0, 0.0],
        "repulsion": [
            [25.0, 0.0],
            [0.0, 25.0]
        ]
   }, {
        "type": "LineString",
        "coordinates": [
           [101.0, 0.0],
           [102.0, 1.0]
        ],
        "repulsion": [
            [10.0, 0.0],
            [0.0, 10.0]
        ]
   }]
}
\end{verbatim}
\end{small}

\vspace{12pt}

\end{document}